\documentclass{article}

\usepackage{PRIMEarxiv}

\usepackage[utf8]{inputenc} 
\usepackage[T1]{fontenc}    
\usepackage{hyperref}       
\usepackage{url}            
\usepackage{booktabs}       
\usepackage{amsfonts}       
\usepackage{nicefrac}       
\usepackage{microtype}      
\usepackage{lipsum}
\usepackage{makecell}
\usepackage{fancyhdr}       
\usepackage{graphicx}       
\usepackage{amsmath}
\graphicspath{{media/}}     
\usepackage{graphicx}%
\usepackage{multirow}%
\usepackage{amsmath,amssymb,amsfonts}%
\usepackage{amsthm}%
\usepackage{mathrsfs}%
\usepackage{threeparttable}
\usepackage[title]{appendix}%
\usepackage{xcolor}%
\usepackage[ruled,linesnumbered]{algorithm2e}
\usepackage{textcomp}%
\usepackage{manyfoot}%
\usepackage{algpseudocode}%
\usepackage{listings}%
\usepackage{array}
\usepackage{tablefootnote}
\title{LSF-IDM: Automotive Intrusion Detection Model with Lightweight Attribution and Semantic Fusion
}

\author{
  Pengzhou Cheng\\
  College of Cyber Science and Engineering  \\
  Shanghai Jiao Tong University \\
  Shanghai, China\\
  \texttt{pengzhouchengai@gmail.com} \\
   \And
  Lei Hua\thanks{\textit{The corresponding author.}} \\
  College of Automotive and Transportation Engineering\\
  Jiangsu University of Technology \\
  Changzhou, China\\
  \texttt{Lei.Hua@cranfield.ac.uk} \\
  \AND
  Haobin Jiang \\
  The Automotive Engineering Research Institute \\
  Jiangsu University \\
  Zhenjiang, China \\
  \texttt{jianghb@ujs.edu.cn} \\
  \And
  Gongshen Liu \\
  College of Cyber Science and Engineering \\
  Shanghai Jiao Tong University \\
  Shanghai, China\\
  \texttt{lgshen@sjtu.edu.cn} \\
}

\begin{document}
\maketitle

\begin{abstract}
Autonomous vehicles (AVs) are more vulnerable to network attacks due to the high connectivity and diverse communication modes between vehicles and external networks. Deep learning-based Intrusion detection, an effective method for detecting network attacks, can provide functional safety as well as a real-time communication guarantee for vehicles, thereby being widely used for AVs. Existing works well for cyber-attacks such as simple-mode but become a higher false alarm with a resource-limited environment required when the attack is concealed within a contextual feature. In this paper, we present a novel automotive intrusion detection model with lightweight attribution and semantic fusion, named LSF-IDM. Our motivation is based on the observation that, when injected the malicious packets to the in-vehicle networks (IVNs), the packet log presents a strict order of context feature because of the periodicity and broadcast nature of the CAN bus. Therefore, this model first captures the context as the semantic feature of messages by the BERT language framework. Thereafter, the lightweight model (e.g., BiLSTM) learns the fused feature from an input packet's classification and its output distribution in BERT based on knowledge distillation. Experiment results demonstrate the effectiveness of our methods in defending against several representative attacks from IVNs. We also perform the difference analysis of the proposed method with lightweight models and Bert to attain a deeper understanding of how the model balance detection performance and model complexity.
\end{abstract}


\section{Introduction}\label{sec1}
\begin{figure}[h]
    \centering
    \includegraphics[width=0.7\linewidth]{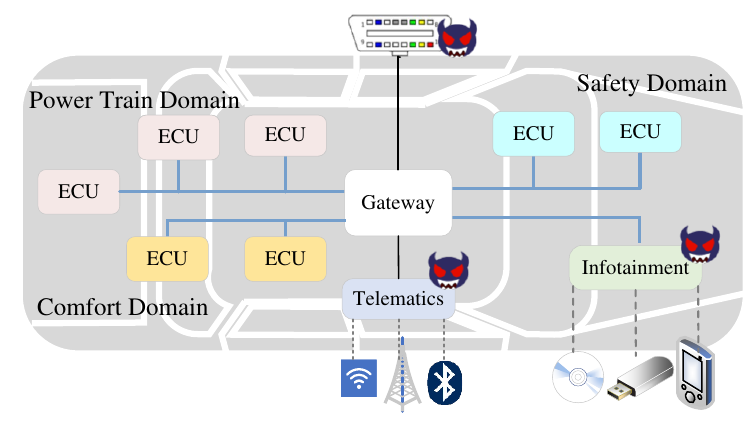}
    \caption{A CAN bus network exploited by attackers through different physical and wireless interfaces}
    \label{fig1}
\end{figure}

Modern vehicles are different from traditional ones as they are fitted with multiple electronic control units (ECUs). These ECUs manage various functions of the vehicle through In-Vehicle Networks (IVNs) like Controller Area Network (CAN)~\cite{song2021self}. In order to improve the travel experience for both drivers and passengers~\cite{seo2018gids, wu2019survey, alkhatib2022can}, more interfaces are open for external networks. Despite the rapid development and great success of IVNs, there have been increasing security concerns about deploying them in real-world applications. The high connectivity between the in-vehicle and the external networks paves the way for hackers to exploit security vulnerabilities in IVN as the weak access control, no confirmation, and no encryption design of the CAN bus~\cite{wu2019survey, javed2021canintelliids, lee2017otids}, as shown in Figure.~\ref{fig1}.

Cyber-attack by injecting malicious packets is one typical security threat~\cite{liu2017vehicle, marchetti2018read}. This idea is that attackers insert well-designed packets into the CAN bus and then hijack the pre-defined protocol (e.g., arbitration mechanism, broadcasting messages to maintain vehicle status and message periodicity)~\cite{song2020vehicle}. It may be fatal if the attacker can control physical subsystems like the engine, steering wheel, and braking system~\cite{nasreldin2021vehicle, abdel2021federated}. Thus, the security vulnerabilities of IVNs are not only an information security or privacy protection issue but also a functional safety-related concern. To protect against various types of cyber-attack, security measures for IVNs involve using encryption and authentication technologies in Cryptography to maintain the confidentiality and integrity of IVNs transmitted messages~\cite{wang2017hardware, van2017vulcan}. Additionally, firewall technology is utilized to separate potential attack interfaces from IVNs~\cite{macher2018signal, macher2017automotive}, and an Intrusion Detection System (IDS) is deployed in AVs to detect attacks on vehicles. However, AVs have numerous attack interfaces, limited computing and storage resources, and high real-time and reliability requirements. Traditional active security protection strategies such as encryption algorithms, identity authentication, and access control are not applicable~\cite{erlacher2020high}. IDS is an efficient protection mechanism used to detect cyber-attacks on automobiles, which can simultaneously provide functional safety as well as a real-time communication guarantee for vehicles.

Initially introduced in the field of traditional methods such as specification-based~\cite{olufowobi2019saiducant, jin2021signature}, statistics-based~\cite{young2019automotive, he2021vehicle, xun2021vehicleeids}, and physical fingerprint-based~\cite{foruhandeh2019simple, kneib2020easi, schell2020valid} to identify intrusion messages but has relatively worse performance. The security research community has recently focused on using deep learning (DL) methods to create efficient IDSs~\cite{erlacher2020high,wu2022rtids, yao2023cnn, al2019intrusion}. These methods model the traffic behavior to identify whether is abnormal or not. However, existing works rely on how to extract valuable features through increasingly complex model structures~\cite{song2020vehicle, lo2022hybrid, cheng2022stc, yue2021ensemble}. Some work focuses on the context feature of transmitted messages as decision-making so as that explore the probability of using pre-trained language models (e.g., BERT and GPT-2)~\cite{alkhatib2022can, nam2021intrusion}. Although these strategies obtain outperformed, it is obviously ineffective to deploy them in a resource-limited environment that IVNs. Another issue is the high false alarm rate (FAR) and a performance bottleneck in a lightweight model~\cite{kang2016intrusion, pawelec2019towards, qin2021application}. 

To address the aforementioned challenges, we present LSF-IDM, an automotive \textbf{I}ntrusion \textbf{D}etection \textbf{M}odel with \textbf{L}ightweight attribution and \textbf{S}emantic \textbf{F}usion. The proposed IDS involves using a pre-trained language model as a teacher model to extract complex contextual semantic features, which can realize valuable feature modeling. Then, we use a lightweight model as a student model to inherit the above knowledge, meanwhile to integrate it into its output probability distribution based on knowledge distillation. Finally, the lSF-IDM can detect abnormal traffic through fused features controlled with a hyper-parameter, by which the bottleneck problem in balancing detection performance and model complexity has been solved. Our contributions are summarized as follows.
\begin{itemize}
    \item We develop a novel lightweight IDM based on semantic fused, which features an approximately pre-trained language model that makes its extracted valuable feature transfer to a lightweight model.
    \item To achieve more effective feature fused, we exploit knowledge distillation, in which the Bidirectional Encoder Representations from the Transformers model (BERT) as the teacher model capture context semantic features and the Bi-directional Long Short-Term Memory (BiLSTM) and Deep neural network (DNN) as student models to identify intrusion and prove the generalization performance.
    \item We evaluate the LSF-IDM on a benchmark car hacking dataset. The experimental results demonstrate the effectiveness of the LSF-IDM in improving detection accuracy and false alarm rate in comparison to traditional machine learning methods. Importantly, an obstacle to practice use in resource-limited is addressed. 
\end{itemize}

The rest of this paper is organized as follows: A comprehensive review of previous works is presented in Section~\ref{sec2}. Section~\ref{sec3} outlines the preliminaries of IVN and the proposed method. Section~\ref{sec4} presents the design of the proposed LSF-IDM in detail. In Section~\ref{sec5}, the setting, results, and analysis of the experiment evaluated are introduced. Finally, we conclude this study and provide some perspectives in Section~\ref{sec6}.

\section{Related Work}\label{sec2}
The DL-based detection methods have been a major research interest in IVN-IDS due to their outstanding performance in the detection accuracy for attacks~\cite{nasreldin2021vehicle}. Recently, several works introduce complex models with strong feature extraction capabilities to design an appropriate feature set to accurately characterize CAN traffic. Seo \textit{et al.}~\cite{seo2018gids} proposed an intrusion detection system based on a generative adversarial network (GAN). The method learned the frequency characteristics of CAN IDs sequentially by training two discriminators. However, it is difficult to train a stable GAN to detect various attacks. Song \textit{et al.}~\cite{song2020vehicle} improved the Inception-ResNet model by proposing the construction of a $29 \times 29$ message matrix in a deep convolutional network. Although their method has a low false-negative and error rate, the introduced model is still a deep structure with high complexity. There are many works focused on extracting the spatial-temporal feature to further optimize performance. 

Lo \textit{et al.}~\cite{lo2022hybrid} used spatial and temporal features for developing an accurate IVN-IDS. The spatio-temporal feature is generated sequentially using a convolutional neural network (CNN) and long short-term memory (LSTM) network. In contrast, Cheng \textit{et al.}~\cite{cheng2022stc} generated spatio-temporal attention features by fetching spatio-temporal features in parallel with CNN and LSTM and introducing an attention module on top of that. This largely makes the model more attentive to important traffic variations. Javed \textit{et al.}~\cite{javed2021canintelliids} combined a convolutional neural network (CNN) and the attention-based Gated Recurrent Unit (GRU) model to propose a novel method named CANintelliIDS. This can detect single intrusion attacks as well as mixed intrusion attacks on a CAN bus. Yue \textit{et al.}~\cite{yue2021ensemble} proposes integrated spatio-temporal feature encoders that extract features in parallel using various variants of LSTM and CNN. The fused features are generated by voting. In order to reduce the model complexity, Tariq \textit{et al.}~\cite{tariq2020cantransfer} preprocesses the in-vehicle telegrams into a two-dimensional spatio-temporal form so as to design the feature encoder directly based on the convLSTM component. Although these features are effective in identifying traffic, they both take advantage of the deep network structure, which does not allow for online detection in resource-constrained environments.

The lightweight model is a necessary requirement for IVN-IDS. Kang \textit{et al.}.~\cite{kang2016intrusion} first applied DNN to IVN-IDS analyzing normal and abnormal packets, which realized a real-time detection model with a 97.8\% detection accuracy on simulation attacks. Taylor \textit{et al.}~\cite{taylor2016anomaly} proposed an IDS constructed using the data field values under the binary of CAN messages based on LSTM. The method reduces the error rate of predicting the n+1st message category through continuous optimization.  Pawelec \textit{et al.}~\cite{pawelec2019towards} extended Taylor's scheme and verified that the proposed method is very promising for data that depends on time series or concurrent inputs. Qin \textit{et al.}~\cite{qin2021application} proposed to input the hexadecimal and binary data domains into LSTM, and with the improved optimization function, the model predicted different CAN IDs separately and obtained a high accuracy. However, all of these methods present a performance bottleneck in terms of identifying malicious messages. It is also straightforward evidence that rich and meaningful features (e.g., spatial, temporal, and context features) are the key to improving model detection.

Recently, the pre-trained language model has an outstanding performance on a variety of downstream tasks in the natural language process (NLP). Due to its powerful modeling of sequences, Alkhatib \textit{et al.}~\cite{alkhatib2022can} first applied BERT in IVN-IDS. The designed CAN-BERT method can learn the context semantic features of CAN ID sequences. Nam \textit{et al.}~\cite{nam2021intrusion} regarded the CAN ID sequence as a sentence with several words and adopt the Generative Pretrained Transformer (GPT) model to learn the normal pattern of each sequence. This method can detect attacks that include a very small number of attack IDs. Although the context feature is significant for IVN-IDS, these advanced models in the research area of NLP are that their large amounts of parameters cannot meet the resource-limited IVN. As a result, researching an intrusion detection model that has high detection capabilities for malicious attacks and can be applied to resource-constrained vehicle environments is vital for ensuring the security of AVs. The comparison between previous works is presented in Table~\ref{tab1}.

\begin{table*}[htbp]
\caption{A comparison of our work with other intrusion detection works in feature extraction capabilities.}\label{tab1}%
\normalsize
\renewcommand\arraystretch{1.3}
\resizebox{\linewidth}{!}{
\begin{tabular}{llccccc}
\toprule
Works& Focus  & \makecell[c]{Semantic \\ feature }  & \makecell[c]{Temporal\\ feature}  & \makecell[c]{Spatial \\ feature} & \makecell[c]{Feature \\ fused} & \makecell[c]{Real-time \\ detection} \\ \midrule
Seo \textit{et al.}~\cite{seo2018gids} & Double discriminator with GAN &  &  & \checkmark & &\\
Song \textit{et al.}~\cite{song2021self}  & Inception-ResNet  &  &  & \checkmark & &\\
lo \textit{et al.}~\cite{lo2022hybrid} & a hybrid model with CNN and LSTM sequentially  & & \checkmark & \checkmark & \checkmark & \\
Cheng \textit{et al.}~\cite{cheng2022stc} & CNN and LSTM with attention mechanism & & \checkmark &  \checkmark& \checkmark & \\
Yue \textit{et al.}~\cite{yue2021ensemble} & Ensemble models with CNN and LSTM &  & \checkmark & \checkmark &\checkmark & \\
Javd \textit{et al.}~\cite{javed2021canintelliids} & CNN and LSTM with attention mechanism & & \checkmark & \checkmark & \checkmark & \\
Tariq \textit{et al.}~\cite{tariq2020cantransfer} & CovLSTM & &\checkmark &\checkmark &\checkmark & \\
Kang \textit{et al.}~\cite{kang2016intrusion} & DNN & & & & & \checkmark \\
Taylor \textit{et al.}~\cite{taylor2016anomaly} & LSTM & &\checkmark & & &\checkmark\\
Pawelec \textit{et al.}~\cite{pawelec2019towards} & 3-layer LSTM & &\checkmark & & &\checkmark \\
Qin \textit{et al.}~\cite{qin2021application} & LSTM in two format & & \checkmark & & & \checkmark\\
Alkhatib \textit{et al.}~\cite{alkhatib2022can} & BERT & \checkmark & & & &\\
Nam \textit{et al.}~\cite{nam2021intrusion} & GPT & \checkmark & & & & \\
Ours & Knowledge distillation, BERT & \checkmark & \checkmark & \checkmark & \checkmark & \checkmark \\
\bottomrule
\end{tabular}}
\end{table*}

\section{Overview of In-vehicle Security}\label{sec3}
In this section, we first introduce the background and nature of the CAN bus and its security mechanism. And then the attack model corresponding to our model is presented. This section describes the background of models used in this work theoretically as well.

\subsection{Controller area network}
The Controller Area Network (CAN) bus, a  broadcast-based networking protocol, was built by Bosch in 1985~\cite{seo2018gids}. It has been widely used for in-vehicle communication due to the low prices, size and functionality, and the efficient and stable connection among ECUs~\cite{tariq2020cantransfer}. 

\begin{figure}[t]
    \centering
    \includegraphics[width=1\linewidth]{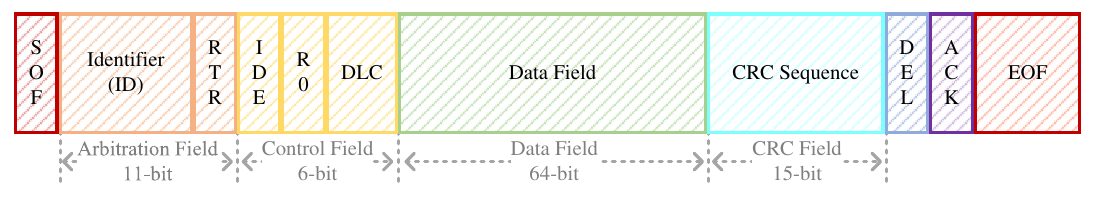}
    \caption{CAN bus standard frame format}
    \label{fig2}
\end{figure}

As shown in Fig.~\ref{fig2}, the standard frame format of the CAN bus consists of Start Of Frame (SOF), IDentifier (ID), Remote frame transmission field (RTR), control field, data field, cyclic redundancy check (CRC), DELimiters (DEL) and ACKnowledgment fields (ACK), and End Of Frame (EOF) fields. Among them, the ID of the arbitration field is to avoid contention by priority mechanism when all ECUs transmit the message simultaneously. This is also mainly a feature used to detect intrusion behaviors as this mechanism provides an opportunity for attackers to launch Denial of Service (DoS) and flooding attacks~\cite{alkhatib2022can}. Due to the characteristics of CAN such as the broadcast mechanism, without authentication and encryption, the attacker can broadcast spoofed CAN messages, eavesdrop on all the CAN traffic, and collect detailed information about it, resulting in fuzzing and spoofing attacks.

\subsection{Attack model}
Since the weak security mechanism of IVN can be confronted with various malicious attacks, as shown in Fig.~\ref{fig3}. In this work, an assumption is determined that the attacker has capable of compromising the ECUs either remotely via wireless interfaces (e.g., the telematic port~\cite{miller2015remote}) or physically (e.g., via OBD-II~\cite{seo2018gids}). Thus, the IVN is susceptible to Denial-of-Service (DoS), Fuzzy, RPM/GEAR, and replay attacks, conducted as follows:

\begin{figure}[t]
    \centering
    \includegraphics[width=1\linewidth]{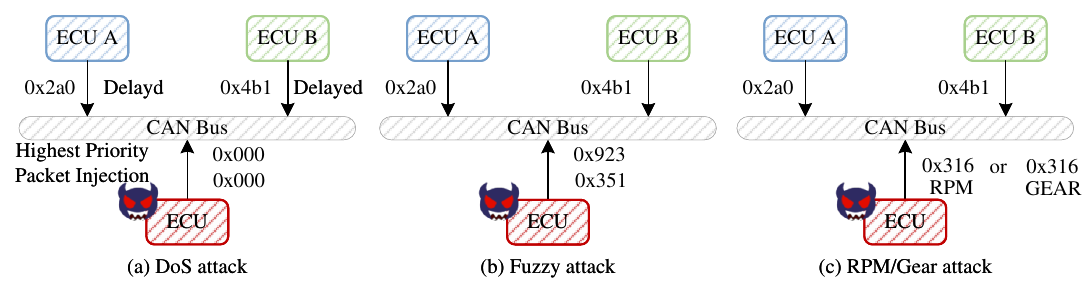}
    \caption{Illustration of the injection process of utilized in‐vehicle intrusion data set}
    \label{fig3}
\end{figure}

\begin{enumerate}
    \item DoS attack: DoS attack, related to availability, is the most common intrusion behavior in IVN as the priority mechanism of CAN bus~\cite{tariq2020cantransfer}. Taking advantage of it, the attackers are able to launch DoS attacks by constantly sending many CAN messages with high priority in a short cycle~\cite{seo2018gids}. 
    \item Fuzzy attack: Attackers can inject CAN messages using spoofed random ID and DATA into the CAN bus~\cite{tariq2020cantransfer}.
    \item RPM/GEAR attack: RPM/GEAR is to inject messages with certain CAN IDs related to RPM/GEAR information~\cite{song2021self}.
\end{enumerate}

\subsection{Models}
\textbf{Pre-trained language model.} Language models aim to learn a probability distribution for representing combinations of words and sentences in natural language texts.
The probability of a sequence of $m$ words $\{w_1, w_2, \cdots, w_n\}$ is denoted as $P\{w_1, w_2, \cdots, w_n\}$~\cite{li2021hidden}. To calculate it, the problem is decomposed with the chain rule of probability:
\begin{equation}
    \begin{aligned}
    P(w_1,\cdots,w_m)&=P(w_1)P(w_2|w_1)P(w_3|w_1,w_2)\cdots P(w_m|w_1,\cdots,w_{m-1})\\
    &=\prod_{i=1}^mP(w_i|w_1,\cdots,w_{i-1}).
    \end{aligned}
\end{equation}

Recently, there has been an increase in the use of pre-trained neural networks in NLP (e.g., BERT~\cite{devlin2018bert}, GPT-2~\cite{radford2019language}). These pre-trained neural networks learn better-contextualized representation and can be applied to several downstream tasks by fine-tuning.

\noindent
\textbf{Deep neural network.} The deep neural network (DNN) $\mathcal{F}(\cdot)$ consists of a sequence of layers $(f_1,f_2,\cdots,f_n)$, where each layer $f_i$ is a differentiable transformation function. Given input $x\in \mathbb{R}^{m\times n}$, the output of the neural network $\mathcal{F}$ is calculated by:
\begin{equation}
    \mathcal{F}(x;\theta)=f_n(f_{n-1}(\ldots(f_2(f_1(x)))))
\end{equation}
where $\theta$ is the parameters of the model. In this work, the model is used to identify the malicious traffic, i.e., a classification task. Suppose there are $c$ different classes, the output of the model $\mathcal{F}(x,\theta) \in \mathbb{R}^c$, where the output distribution can represent as $\mathcal{F}(x, \theta)_o = \{o_1, o_2, \cdots, o_c\}$ so that the maximum probability is the final class. In order to achieve a non-linear fit, some activation functions are introduced such as sigmoid, tanh, Rectified Linear Unit (ReLU), and softmax function.

\noindent
\textbf{LSTM.}
Initially, the Recurrent neural network (RNN) presents feedback loops that impact the model's output using the same input values and network state. To address the vanishing gradient problem from RNN, the Long-Short Term Memory (LSTM) model is proposed by adding gating mechanisms, including input gates, output gates, and forget gates. When the input sequence $X = \{x_1, x_2, \cdots, x_t\}$, the LSTM can compute the output sequence $Y=\{y_1, y_2, \cdots, y_t\}$, where the activation states of neurons $c_t$ and memory units $m_t$ at time $t$. The memory cells in the LSTM are iteratively updated as follows:
\begin{equation}
    \begin{gathered}
        i_t =\sigma(W_i\cdot[h_{t-1},x_t]+b_i), \\
        o_t =\sigma(W_o\cdot[h_{t-1},x_t]+b_o), \\
        f_{t}=\sigma\left(W_{f}\cdot[h_{t-1},x_{t}]+b_{f}\right), \\
        \tilde{C}_{t}=\mathrm{tanh}(W_{c}\cdot[h_{t-1},x_{t}]+b_{c}), \\
        C_{t}=f_{t}\odot C_{t-1}+i_{t}\odot\tilde{C}_{t}, \\
        h_{t}=o_{t}\odot\tanh{(C_{t})}, \\
        y_{t}=W_{y}h_{t}+b_{y}, 
    \end{gathered}
\end{equation}
where the $i_t, o_t, f_t$ donate the input, output, and forget of the gate units at time $t$, respectively. The $W$ and $b$ represent the corresponding weight and bias. The $y_t$ is the final prediction class of a specific task. Actually, all of these gate units decide what information is important and reserved by the activation function.

\section{Methodology}\label{sec4}
In this section, we present LSF-IDM that can classify IVN traffic by exploiting a valuable fused feature. We first describe the data preprocessing step. Then, we present the LSF-IDM framework in detail, including the teacher-student paradigm and knowledge distillation.

\subsection{Data preprocessing}\label{sec4.1}
Data preprocessing is a crucial step for the DNN module as it is sensitive to the quality of training data. However, the CAN traffic has many unexpected problems when the log is saved, such as incomplete, and inconsistent. In this work, we use the car hacking for intrusion detection dataset from the hacking and countermeasure research lab (HCRL)\footnote{https://ocslab.hksecurity.net/Datasets/car-hacking-dataset}. The dataset was created by recording CAN packets through the OBD-II port of a vehicle during various CAN injection attacks, including DoS attack, fuzzing attack, RPM attack, and GEAR attack. Table~\ref{tab2} presents the attack types and size of this dataset, with the number of samples for the training and test dataset. 

\begin{table}[t]
\centering
\caption{HCRL Dataset Characteristics}\label{tab2}%
\begin{tabular}{p{2cm}>{\centering\arraybackslash}m{1in}cc}
\toprule
Type & Raw Data  & \#Number of Training Samples & \#Number of Test Samples\\
\midrule
Normal    & 14,037,293   &  9,826,105&	4,211,188  \\
DoS	&587,521	&411,265	&176,256  \\
Fuzzy 	&491,847	&344,293	&147,554  \\
RPM	&654,897&	458,428&	196,469 \\
Gear	&597,252	&418,076	&179,176 \\ 
\bottomrule
\end{tabular}
\end{table}

\begin{figure}[h]
    \centering
    \includegraphics[width=0.7\linewidth]{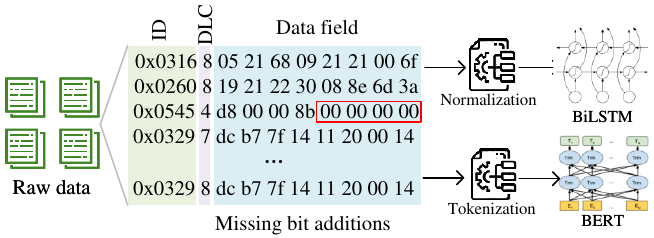}
    \caption{Data preprocessing}
    \label{fig4}
\end{figure}

In this dataset, the feature of each message involves the timestamp, CAN ID, Data Length Code (DLC), data fields (data[0]-data[7], totaling 8 bytes), and a flag bit (Flag) used to mark the class of the CAN message. As shown in Fig.~\ref{fig4}, the data preprocessing module can convert the captured data to a uniform scale in order to meet the input of deep learning. 
Specifically, the data field is divided into eight characteristic fields, where each field contains two hexadecimal values. In case any bytes are missing in the data field, the value "00" is used to fill them up. Moreover, all hexadecimal values within the ID and Data domain are converted into decimal values. Note that we only consider the field of ID, DLC, and Data as the features because of the important correlations with the representative attacks.

Subsequently, data preprocessing requires independent implementations corresponding to the student model and the teacher model. In terms of the BERT model, CAN messages need to be transformed into word lists by Tokenize operation and then Embedding is performed on them, while for BiLSTM, the data needs to be normalized. Formally, the training data $\mathbb{D}$ is a pair-wise instance of a dictionary that includes the lightweight model input $(X_i, y_i) = \{x_1, x_2, \cdots, x_n; y_i\}^{1\times10}$ corresponding to the teach model input $(X_i, y_i) = \{token_1, token_2, \cdots, token_n; y_i\}^{1\times \mathrm{max\_length}}$. The $\mathrm{max\_length}$ is a hyper-parameter from BERT.

\subsection{LSF-IDM}
\begin{figure}[t]
    \centering
    \includegraphics[width=1\linewidth]{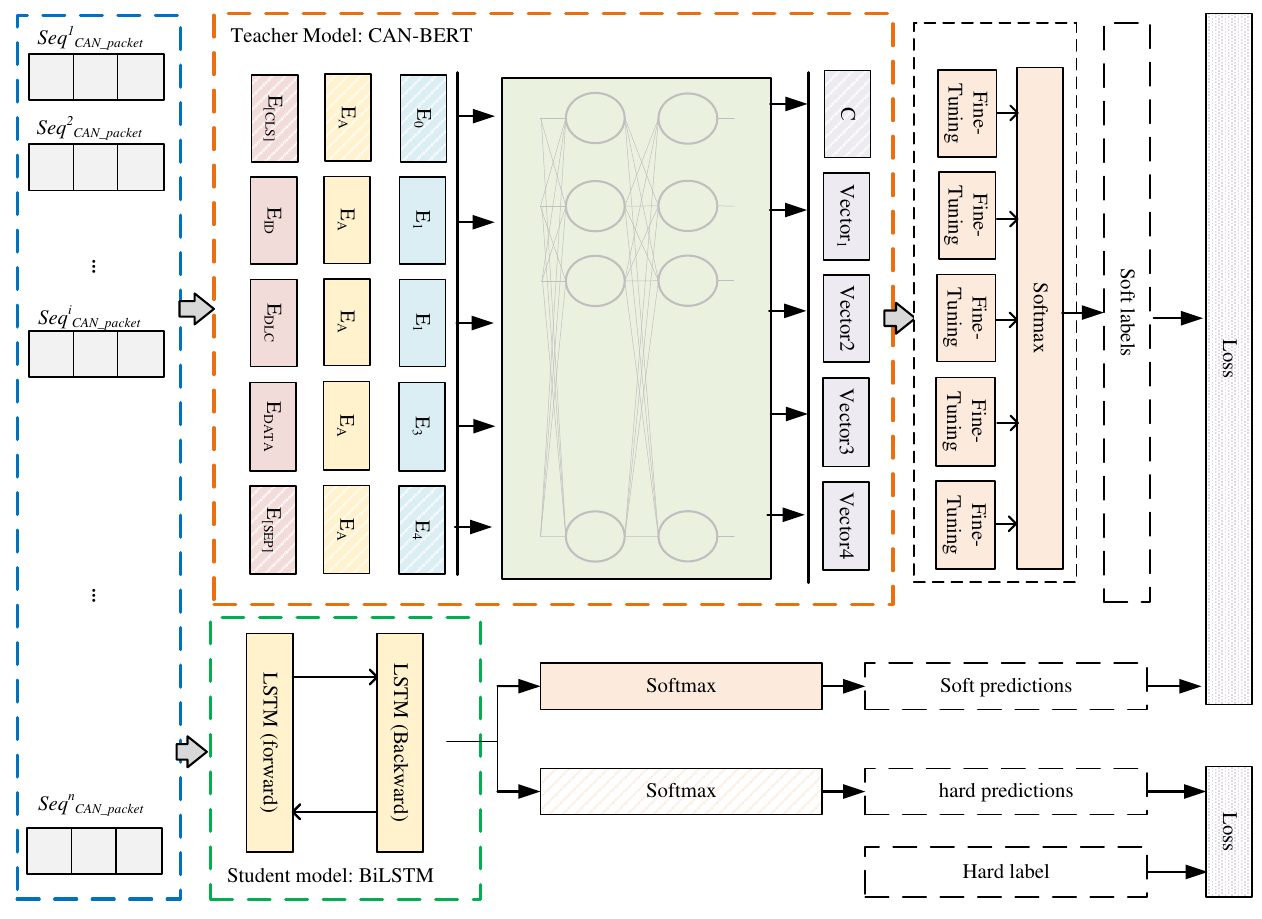}
    \caption{Proposed LSF-IDM pipeline}
    \label{fig5}
\end{figure}

In state-of-the-art IDSs, frame detection usually pursues either outstanding performance via complex model structure or real-time by smaller parameters.  It is certain that a balance needs to be achieved between detection performance and low-resource requirements. To this end, we present a semantic fusion-based malicious attack detection model, named LSF-IDM. The Fig.~\ref{fig5} shows the overall workflow. Given the training dataset and a teacher-student model, the LSF-IDM pipeline consists of three steps to determine whether the CAN bus is intrusion:
\begin{enumerate}
    \item \textbf{Context semantic feature generation.} The teacher model first learns the contextual semantic features of CAN messages through a pre-trained language model (our method is based on BERT). This can capture significant features and then improve the detection performance due to its high model complexity and a large number of parameters (Transformer blocks=12, Hidden size=768, Total parameters=110M).
    \item \textbf{Lightweight model training.} We use a lightweight model (e.g., BiLSTM or DNN) in this step. We design a feature extractor to get representation vectors of the lightweight model.
    \item \textbf{Feature fused.} Given a teacher model, the lightweight model as a student model also learns its knowledge. The fused feature is composed of representation vectors from a teacher-student model based on knowledge distillation. Then, we train a classifier to detect anomalies using the fused feature.
\end{enumerate}

 As a result, the proposed can be directly applied to resource-constrained in-vehicle network environments. The whole LSF-IDM framework is presented in the Algorithm~\ref{algo1}. In the following, we first introduce our approach to context semantic feature generation- using BERT  in Section~\ref{4.2.1}. Then, we introduce the way to perform the lightweight model training in Section~\ref{4.2.2}. Finally, the feature fused-knowledge distillation is present in Section~\ref{4.2.3}. As a result, employing it as the student model of LSF-IDM and using knowledge distillation to simulate the output distribution of the CAN-BERT model while learning CAN message classification. 
 
\begin{algorithm}[h]
\caption{LSF-IDM}\label{algo1}

\KwIn {$(X_i, y_i) \in \mathbb{D}_{train}$}
\KwOut {Detection result}
\tcc{Context-semantic feature generation and teacher model training}
\While{the teacher model is not convergence}{
    \For{$ (X_i, y_i)_{teacher} \in \mathbb{D}_{train}$}{
        $V_i \leftarrow \mathrm{BERT}(X_i)$\; 
        $\hat{y_i} \leftarrow \mathrm{MLP}(V_i)$\;   
        \tcc{$\hat{y_i}$ is context-semantic features}
        $loss \leftarrow \mathcal{L}_{CE}(y_i, \hat{y_i})$\;  
        $loss.backward()$\; 
    }
}
\tcc{Feature fused and lightweight model training}
\While{the student model is not convergence}{
    \For{$ (X_i, y_i)_{student}, (X_i, y_i)_{teacher} \in \mathbb{D}_{train}$}{
        \tcc{generate context-semantic feature}
        $\hat{y}_i^{teacher} \leftarrow \mathrm{BERTforClassification}(X_i^{teacher})$\; 
        \tcc{F is a lightweight model}
        $V_i^{student} \leftarrow \mathcal{F}(X_i^{student})$\;
        $\hat{y}_i^{student} \leftarrow \mathrm{MLP}(V_i^{student})$\;    
        $loss \leftarrow \alpha\mathcal{L}_{CE}(y_i, \hat{y_i}) + (1-\alpha)\mathcal{L}_{KD}(\hat{y}_i^{student}, \hat{y}_i^{teacher})$\;  
        $loss.backward()$\; 
    }
}
\tcc{Intrusion detection evaluation}
\For{$ (X_i, y_i)_{student} \in \mathbb{D}_{train}$}{
    $V_i^{student} \leftarrow \mathcal{F}(X_i^{student})$\;
    $\hat{y}_i^{student} \leftarrow \mathrm{MLP}(V_i^{student})$\;
    $\mathrm{Acc} += \sigma(\hat{y}_i^{student})_{max-probability} == y_i$
}
\end{algorithm}

\subsubsection{Context semantic feature generation}\label{4.2.1}
We hope to generate valuable context-semantic features so that the classifier can accurately identify attacks. This is essential because an observation that the attacker injection attacks may cause a strict order in the message log. 

To this end, we propose the use of a pre-trained language model-BERT\footnote{https://huggingface.co/bert-base-uncased}, which pre-trains bidirectional representations from the unlabeled text by joint conditioning on both the left and right context. To our best knowledge, only one study on IVN-IDS employed it detecting anomaly detection, that is~\cite{alkhatib2022can}. However, it is impractical to use the fact that there are more parameters due to BERT directly in IVNs with limited resources. Inspired by it, the first step of our defense pipeline is to generate a context-semantic feature, based on which the defender can exploit to transfer instructive knowledge to a lightweight model. 

Given a preprocessed message $X_i = \{token_1, token_2, \cdots, token_n\} \in \mathbb{R}^{d_x} $ with a ground truth label $y \in \{1,\cdots, c\}$. to BERT, we first compute the token embedding vector by a mapping function $\mathcal{I}$ as follows:
\begin{equation}
    T_i = \mathcal{I}(X_i, \Theta) = W_eX_i + b_i,
\end{equation}
where $\Theta$ is the parameter of the mapping function. The $W_e$ and $b_e$ are the input embedding weight matrix and bias, respectively. Thereafter,
the CAN messages' position is encoded into positional embedding $P_i$ using a sinusoidal function. 

Having the embedding vector $E_i = T_i + P_i$, we will feed it into a stack of $L$ transformer encoders. Without loss of generality, we select $l-th \in L$ layer for the interpretation of contextual semantic feature extraction. Each transformer layer has two sub-layers: a multi-head self-attention module $g$ and a position feed-forward network $z$. The residual connection establishes the information transfer mechanism between the two sub-layers and the layer normalization $f$ improves the model generalization ability. The final output vector is thus calculated as follows:
\begin{equation}
    \begin{aligned}
           M_i &= g(E_i) + f(E_i + g(E_i)), \\
           H_i & = z(M_i) + f(M_i + z(M_i)) 
    \end{aligned}
\end{equation}
where the $M_i$ and $H_i$ are the output of two sub-layer in $l-th$, respectively. 

The multi-head self-attention module $g$ is a crucial attribution for generating the context-semantic feature. To simplify and clarify the problem,  we formally present how to calculate the attention feature of the embedding vector $E_i$ of the CAN message in one head, calculated as follows: 
\begin{equation}
    Attention(Q^l,K^l,V^l)=AV^l = \sigma\left(\frac{Q^lK^{(l)^T}}{\sqrt{d}}\right)V^l,
\end{equation}
where $A \in \mathbb{R}^{n\times n} $ is the attention matrix and n is the dimension of the CAN message. $\sigma(\cdot)$ is the softmax function; $d$ is the dimension of the vectors  $Q^l, K^l$ and $V^l$; and $\sqrt{d}$ can solve the gradient vanishing problem. $Q^l, K^l$ and $V^l$ represent projections of the embedding vector $E_i$ using scaled dot-product attention. 

In fact, the generated features refer to a strong correlation between fields of the CAN message if the query and key have a high attention weight. We believe that this capture of global information is feasible for anomaly detection so that can extract valuable context-semantic feature extraction. The context semantic feature generation corresponds to the algorithm~\ref{algo1} in lines 1-8. Not that the optimizer function is cross-entropy.

\subsubsection{lightweight model training}\label{4.2.2}
The defender will perform the lightweight model training algorithm based on the context-semantic features generated by the teacher model. Our lightweight model consists of two goals: 1) learn the context-semantic distribution from the teacher model; 2) integrate all valuable information from CAN messages and train a classifier to distinguish normal and abnormal traffic. In the following, we will first introduce BiLSTM as our lightweight model because its good sequence modeling ability allows for a better understanding of the knowledge passed on. And then we introduce the DNN as our lightweight model to show the generality of semantic features of context-semantic.

\textit{BiLSTM.}
Compared with Long Short-Term Memory (LSTM), Bi-directional Long Short-Term Memory (BiLSTM) has a strong capability of capturing contextual features so that it can better learn the contextual semantic information of CAN messages. Therefore, we use the BiLSTM which is a combination of a forward LSTM and a backward LSTM as the student model to learn the global information in the BERT model, calculated as follows:
\begin{equation}
    \begin{aligned}
\overrightarrow{hw_t} &=\overrightarrow{LSTM}(w_{t})  \\
\overleftarrow{hw_t} &=\overleftarrow{LSTM}(w_t) \\
hw &=\left[\overrightarrow{hw_t}\circ\overleftarrow{hw_t}\right]\\
\end{aligned}
\end{equation}
where $\overrightarrow{hw_t}$ and $\overleftarrow{hw_t}$ donate the hidden state of forward and backward, respectively. And $hw$ is the output vector of the model.

\textit{DNN.} Meanwhile, we propose to use a two-layer fully connected neural network as another lightweight model. Although it is not able to absorb all the knowledge from the teacher model as well as BiLSTM, we hypothesize that this knowledge also plays a guiding role (as evidenced by the performance variability analysis).

\subsubsection{Feature fused}\label{4.2.3}
Knowledge distillation (KD) can transfer the ``knowledge" of a complex model to a lightweight model, and make its performance close to the complex model, which can solve the problem of balancing performance and model complexity~\cite{hinton2015distilling}. The distillation process is shown in equation~\ref{eq8}, where $Z_i$ is the logits, $ p_i$ is the probability that the input belongs to category $i$, and $T$ is the super-parameter ``temperature". The larger $T$ is the more uniform the distribution of the output probability.

\begin{equation}\label{eq8}
    p_i=\frac{\exp\left(\frac{Z_i}T\right)}{\Sigma_j\exp\left(\frac{Z_j}T\right)}.
\end{equation}

Formally speaking, let $\{(X_i^{teacher}, y_i)\}^m_{i=1}$ donate the training data for BERT, where $X_i: \mathbb{R}^{d_x} \rightarrow \{1, \cdots, c\}$ is the CAN message and $y_i$ is a binary label indicating whether $X_i$ is abnormal ($y_i=1$) or normal ($y_i=0$). We feed it into the convergence teacher model and then obtain the context-semantic distribution $\hat{y}_i^{teacher}$. Similarly, the $\{(X_i^{student}, y_i)\}^m_{i=1}$ is fed into the lightweight model, we also get the output distribution $\hat{y}_i^{student}$. In order to achieve knowledge transfer, we use the loss function $\mathcal{L}_{KD}$, represented as follows:

\begin{equation}
    \mathcal{L}_{KD}=-\frac1K\sum_{i,c}[\hat{y}^{teacher}_{i,c}\cdot\log P^S\left(y_i=c|X_i;\Theta\right)],
\end{equation}
where $c$ is the sample category, $\Theta$ is a parameter of the student model, and $P^S(y_i=c|X_i; \Theta)$ is the probability of belonging to $c$ computed by the student model for the sample $X_i$. Consistent with the teacher mode, the LSF-IDM is trained with the $\mathcal{L}_{CE}$, represented as follows:
\begin{equation}
   \mathcal{L}_{CE}=\frac{1}{N}\sum_{i}-[y\log\hat{y_i}+(1-y)\log{(1-\hat{y})}]
\end{equation}

In summary, all the objective functions in the whole model training are shown in Equation~\ref{eq11}, in which $\alpha$ is the hyper-parameter used to balance the $\mathcal{L}_{KD}$ and $\mathcal{L}_{CE}$. All process is present in Alogrithm~\ref{algo1} in lines 9-17.

\begin{equation}\label{eq11}
    \mathcal{L} =  \underset{\theta}{\arg\min}\sum_{i} \alpha\mathcal{L}_{CE}(y_i^{student}, \hat{y_i^{student}}) + (1-\alpha)\mathcal{L}_{KD}(\hat{y}_i^{student}, \hat{y}_i^{teacher})
\end{equation}

\subsubsection{Target traffic detection}
Let $\Theta^*$ represent the parameter obtained by the LSF-IDM. Given a targeted traffic $X_i$, we only apply the lightweight model, i.e., LSF-IDM with $\Theta^*$ to determine whether it is hijacked or not. Importantly, the time complexity of the proposed model is equivalent to the previously lightweight models~\cite{kang2016intrusion, pawelec2019towards, qin2021application}, and with more sensitive recognition of malicious attacks.

\section{Experiment and Analysis}\label{sec5}
In this section, we will introduce the experiment setting and evaluation metrics. Then, the results of using our pipelines to detect abnormal traffic on the car-hacking dataset are presented. The experimental results are devoted to proving the validity of LSF-IDM and its superiority over traditional lightweight methods by learning context-semantic representation of the in-vehicle network traffic. 

\subsection{Experiment setting}
The proposed LSF-IDM has been implemented in Python language using the Pytorch library for deep learning methods. We conducted experiments on a machine with Intel (R) Xeon(R) Gold 6271C CPU @ 2.60GHz, 32GB RAM, and NVIDIA Tesla T4 (16G). Table~\ref{tab3} presents the experiment setup in detail. 
\begin{table}[t]
\centering
\caption{Experimental environment}\label{tab3}%
\begin{tabular}{@{}lll@{}}
\toprule
& Parameter & Value \\
\midrule
\multirow{3}{*}{Hardware} &CPU    &   Intel(R) Xeon(R) Gold 6271C CPU @ 2.60GHz  \\
&RAM & 64G \\
&Graphics card \& Memory & NVIDIA Tesla T4 (16G) \\ \midrule
\multirow{4}{*}{Software} & Pytorch & 1.12.1 \\
& Cuda & 11.3 \\
& Cudnn & 8.1.0 \\
& Tensorboard & 2.11.2 \\
& Python & 3.9.16 \\
\bottomrule
\end{tabular}
\end{table}
\subsection{Evaluation Metrics}
In order to comprehensively evaluate the performance of the proposed method, we used accuracy (ACC), precision (PRE), recall (REC), and F1-score as performance metrics. We first require to define the confusion matrix with respect to IVN-IDS, as follows:
\begin{itemize}
    \item True Positive (TP): Malicious traffic is detected successfully by the LSF-IDM.
    \item True Negative (TN): Normal traffic is identified as non-intrusion behavior successfully by the LSF-IDM.
    \item False Positive (FP): Normal traffic is identified as intrusion behavior, related to the false alarm.
    \item False Negative (FN): Malicious traffic is detected failure, related to miss alarm.
\end{itemize}

According to the definition of  the confusion matrix, the aforementioned metrics are calculated as follows:
\begin{equation}
    ACC = \frac{TP+TN}{TP+FP+FN+TN}
\end{equation}
\begin{equation}
    PRE = \frac{TP}{TP+FP}
\end{equation}
\begin{equation}
    REC = \frac{TP}{TP+FN}
\end{equation}
\begin{equation}
    F1-score = \frac{2PRE\cdot REC}{PRE+REC}
\end{equation}

where ACC represents the ratio of correctly classified instances and the total number of instances. PRE donates the fraction of target instances predicted as positive that is actually positive, while REC presents the fraction of positive instances that are detected. The F1 score is the harmonic mean of precision and recall and is commonly used to evaluate model performance on unbalanced datasets.

In particular, we exploit the false positive rate (FPR), and false negative rate (FNR) to evaluate the false alarm and miss alarm of the proposed model, as follows:
\begin{equation}
    FPR = \frac{FP}{TN+FP}
\end{equation}
\begin{equation}
    FNR = \frac{FN}{TP+FN}
\end{equation}

\begin{table}[t]
\centering
\caption{Hyperparameter settings}\label{tab4}%
\begin{tabular}{p{3cm}lp{3cm}lp{3cm}l}
\toprule
Model& Parameter & Value \\
\midrule
\multirow{4}{*}{BERT} &Batch size    &   128  \\
&Learning rate & $5\times10^{-5}$ \\
&Epoch & 3 \\
&Max length & 16 \\ \midrule
\multirow{5}{*}{LSF-IDM} & Learning rate & $10^{-5}$ \\
& Epoch & 8 \\
& Batch size & 1024 \\
& Hidden size & 64 \\
& LSTM layer & 2 \\
\bottomrule
\end{tabular}
\end{table}

\subsection{Detection performance}
\begin{figure}[t]
    \centering
    \includegraphics[width=1\linewidth]{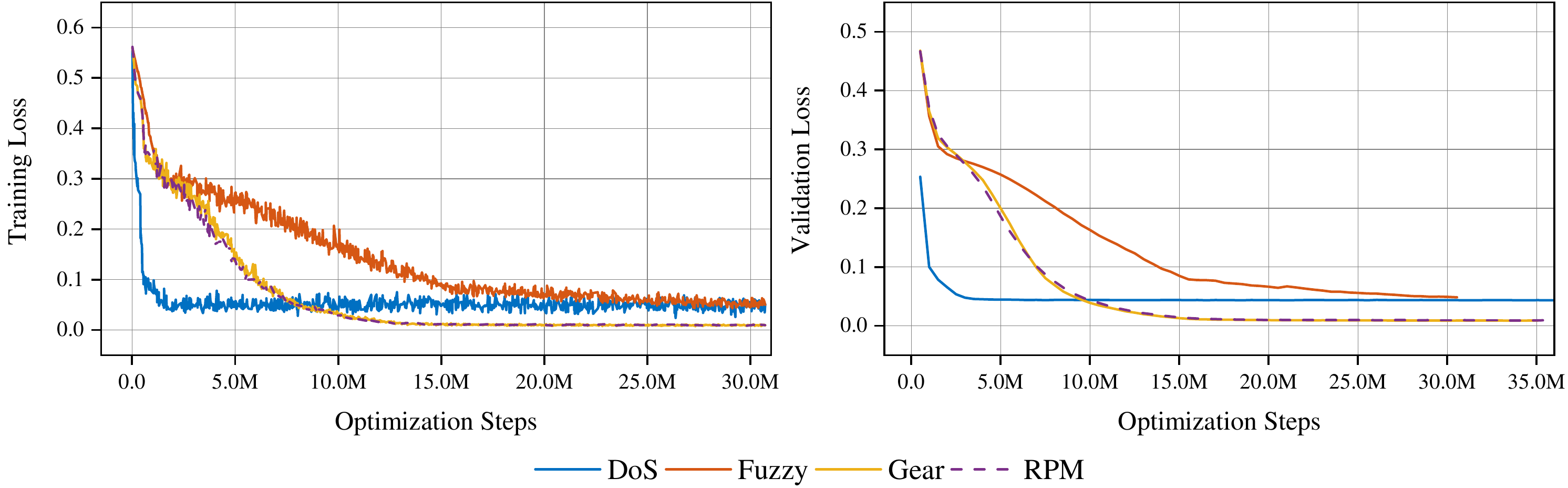}
    \caption{Training and testing performance of LSF-IDM detector for four attack datasets.}
    \label{fig6}
\end{figure}

Hyperparameters are a critical factor in determining the performance of a model. In this study, some parameters are determined before training the LSF-IDM, as shown in Table~\ref{tab4}. In terms of the teacher model, we define the batch size as 128, the learning rate as $5\times 10^{-5}$, the epoch as 3, and the max length as 16, respectively. For the lightweight model, the learning rate is $10^{-5}$, the epoch is 8, the batch size is 1024, the hidden size is 64, and the LSTM layer is 2. Based on it, we trained the LSF-IDM and then show the detection performance. The fig.~\ref{fig6} presents the optimization results of the training phase and validation phase.

Intuitively, the chosen hyperparameters allow the model to significantly enhance inference speed, with the DoS attack having the quickest decline in loss and presenting sub-optimal convergence values. This result can be attributed to its attack principles that force attack frames to be injected into the CAN bus with the highest priority. In contrast, the training of fuzzy attack injected with random data has a maximum loss value, until 30M steps gradually form a convergence state. With the same attack setting, the spoofing attack of Gear/RPM has similar convergence status, achieving optimal performance. Note that trained-well models will be deployed to evaluate detection performance on the testing dataset, respectively. The validation result is consistent with the training phase so that the proposed model can identify various attacks.

\begin{figure}[t]
    \centering
    \includegraphics[width=1\linewidth]{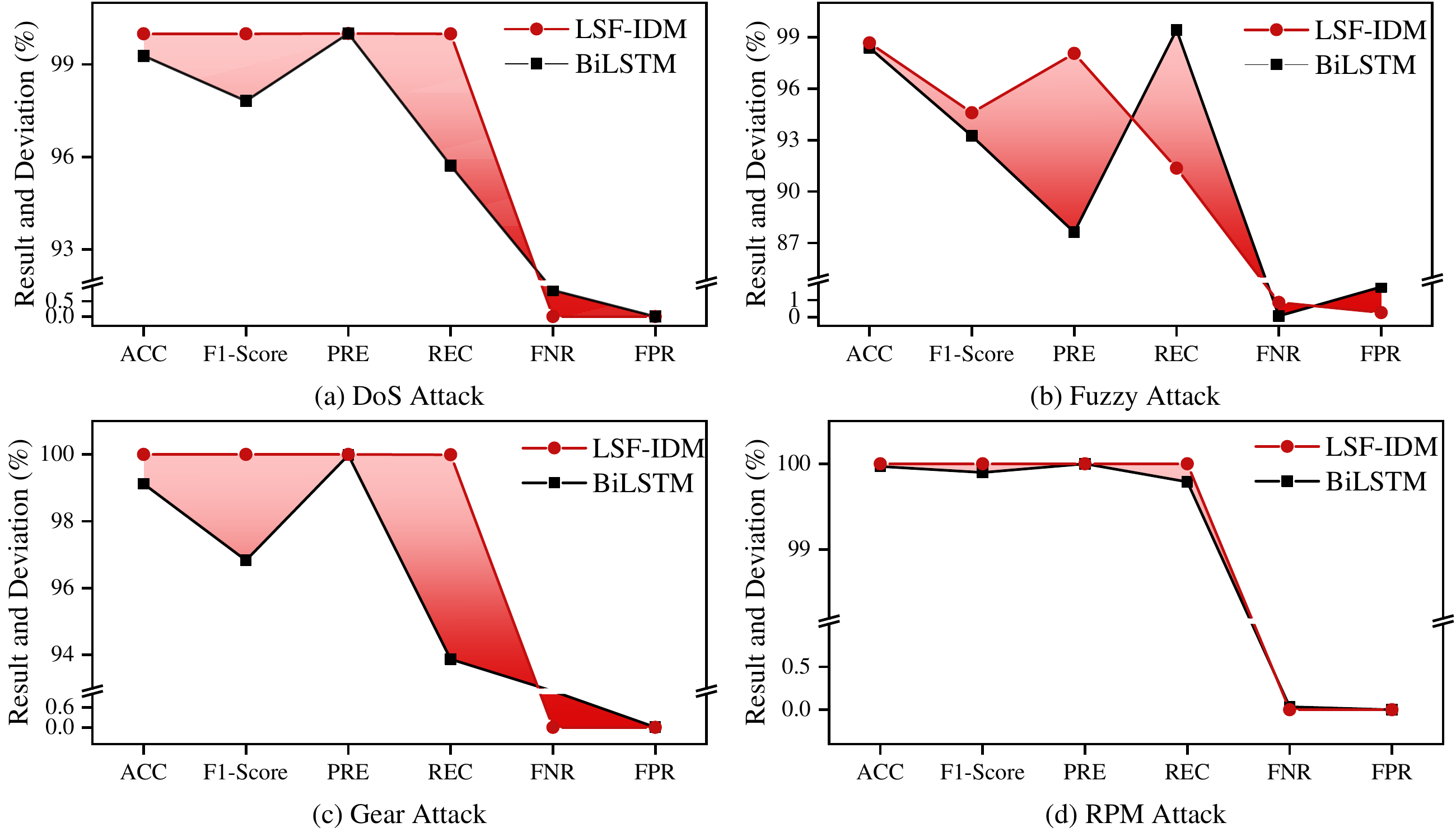}
    \caption{Performance evaluation on each type of attack of the HCRL dataset, where each type from (a) to (d) consists of evaluation performance on six metrics between the lightweight model and the LSF-IDM. In particular, the upper and the bottom deviation of both models are filled with color respectively.}
    \label{fig7}
\end{figure}
In Fig.~\ref{fig7}, we show the evaluation results and deviation between the lightweight model and the LSF-IDM.
Overall, it can be seen that when the lightweight model learns the knowledge that is context-semantic from the teacher model, the performance on the test sets increases for all four attacks. For Dos attacks and spoofing attacks, the proposed model presents comprehensive advantages with outstanding performance and lower false alarms. This result attributes to the focus on context information of the CAN message, because the DoS attack that the highest priority of frames injected in a short period could corrupt this context information while a spoofing attack that modified the data content can generate inconsistency frame with the previous broadcast in the CAN bus. In contrast, it is difficult to detect the fuzzy attack. Despite having a lower recall, our model also maintains an extent ahead of the lightweight model, especially in the F1-score. In terms of false alarm and miss alarm, the proposed model has stable performance, which is crucial for drivers. The results validate our motivation that learning larger models' knowledge to apply in a lightweight model is work through knowledge distillation. It not only breaks the performance bottleneck but also realizes the purpose of excellent detection performance in real-time.

\subsection{Performance difference analysis}
\begin{table}[t]
\caption{Performance difference analysis of LSF-IDM with BERT and BiLSTM }
\label{tab5}
\resizebox{\linewidth}{!}{
\begin{tabular}{llllllll}
\toprule
Attack type                   & Model                                   & ACC/\% & F1-Score/\% & PRE/\% & RE/\% & FNR/\%      & FPR/\%      \\ \hline
\multirow{5}{*}{DoS}   & BERT                                 & 100    & 100         & 100    & 100   & 0           & 0           \\
                        & \multirow{3}{*}{LSF-IDM} & \textcolor{green}{-0.72}  & \textcolor{green}{-0.01}       & \textcolor{green}{0}      & \textcolor{green}{-0.01} & \textcolor{green}{0}          & \textcolor{green}{0}           \\
                       &                                      & 99.28  & 99.99       & 100    & 99.99 & 0           & 0           \\
                       &                                      & \textcolor{red}{+0}     & \textcolor{red}{+2.18}       & \textcolor{red}{0}      & \textcolor{red}{+4.27} & \textcolor{red}{-0.85}       & \textcolor{red}{0}           \\
                       & BiLSTM                               & 99.28  & 97.81       & 100    & 95.72 & 0.85        & 0           \\ \midrule
\multirow{5}{*}{Fuzzy} & BERT                                 & 99.98  & 99.98       & 99.97  & 99.98 & 0.01        & $2.99\times10^{-3}$  \\
                       & \multirow{3}{*}{LSF-IDM} & \textcolor{green}{-1.29}  & \textcolor{green}{-5.24}      & \textcolor{green}{-1.79}  & \textcolor{green}{-8.37} & \textcolor{green}{+0.85}       & \textcolor{green}{+0.25}       \\
                       &                                      & 98.69  & 94.74       & 98.18  & 91.53 & 0.86        & 0.25        \\
                       &                                      & \textcolor{red}{+0.29}  & \textcolor{red}{+1.36}       & \textcolor{red}{+10.43} & \textcolor{red}{-8.06} & \textcolor{red}{+0.79}       & \textcolor{red}{-1.52}       \\
                       & BiLSTM                               & 98.37  & 93.23       & 87.62  & 99.42 & 0.07        & 1.77        \\ \midrule
\multirow{5}{*}{Gear}  & BERT                                 & 100    & 100         & 100    & 100   & 0           & 0           \\
                       & \multirow{3}{*}{LSF-IDM} & \textcolor{green}{0}      & \textcolor{green}{-0.01}           & \textcolor{green}{0}      & \textcolor{green}{0} & \textcolor{green}{0}           & \textcolor{green}{0}           \\
                       &                                      & 100    & 99.99       & 100    & 100   & 0           & 0           \\
                       &                                      & \textcolor{red}{+0.88}  & \textcolor{red}{+3.16}       & \textcolor{red}{+0.02}  & \textcolor{red}{+6.13} & \textcolor{red}{-1.01}       & \textcolor{red}{$-3.53\times10^{-3}$}  \\
                       & BiLSTM                               & 99.12  & 96.83       & 99.98  & 93.87 & 1.01        & $3.61\times10^{-3}$   \\ \midrule
\multirow{5}{*}{RPM}   & BERT                                 & 100    & 100         & 100    & 100   & 0           & 0           \\
                       & \multirow{3}{*}{LSF-IDM} & \textcolor{green}{0}      & \textcolor{green}{0}           & \textcolor{green}{0}      & \textcolor{green}{0}     & \textcolor{green}{$+3.35\times10^{-12}$} & \textcolor{green}{$+1.46\times10^{-13}$} \\
                       &                                      & 100    & 100         & 100    & 100   & $3.35\times10^{-12}$ & $3.35\times10^{-13}$ \\
                       &                                      & \textcolor{red}{+0.03}  & \textcolor{red}{+0.10}       & \textcolor{red}{0}      & \textcolor{red}{+0.21} & \textcolor{red}{-3.4e-4}     & \textcolor{red}{+1.46e-13}   \\
                       & BiLSTM                               & 99.97  & 99.90       & 100    & 99.79 & 0.03        & 0           \\ \bottomrule
\end{tabular}}
\footnotesize{Note: The data in red is the percentage increase in the performance of our model compared to BiLSTM (where FNR and FPR represent the error rate, the smaller the value the better the performance of the model); The data in green shows the percentage decrease in the performance of this model compared to BERT, which has the best performance. BERT has a huge number of parameters that is difficult to be applied to the IVNs. }
\end{table}
To further prove the contribution of the fused feature,  we conduct the performance difference analysis on the LSF-IDM with the lightweight model and the pre-trained language model. Table~\ref{tab5} shows the results of the four attacks, as well as the rise performance relative to the lightweight model and the fall performance of the teacher model. 

The testing for the ACC metric on the LSF-IDM achieves 99.28\% on DoS, 98.69\% on Fuzzy ($0.29\% \uparrow$), and 100\% on spoofing attacks ($0.88\%\uparrow$ and $0.33\%\uparrow$) with respect to the lightweight model. Due to the inability to fully absorb the fusion knowledge, 0.72\% and 1.29\% of the performance is lost on DoS attacks and Fuzzy attacks. In terms of the DoS attack, due to the precision of 100\%, the instructional capacity is reflected in the recall rate (95.72\%$\rightarrow$99.99\% with raising 4.27\%$\uparrow$), while only decreasing 0.01\% from BERT. Moreover, the performance enhancement in the recall is also presented in Gear/RPM attacks with 6.13\% and 0.21\%, which can identify malicious traffic with full knowledge of BERT knowledge. It is worth noting that the F1-score presents dramatic improvement in the above attacks (2.18\%$\uparrow$, 3.16\%$\uparrow$ and 0.1\%$\uparrow$) as the enhancement in precision and recall. Differently, the LSF-IDM does not seem to be able to learn to recall capabilities on fuzzy attacks from BERT, but precision is improved significantly (10.43\%$\uparrow$). This reason is that the proposed model prefers to capture the true anomaly. Thus, the overall performance in F1-score achieves 94.74\%.

Interestingly, the proposed model controls well on the FAR and the FNR. We observe that the error rate is hardly and approximately 0. Although having 0.86\% and 0.25\% in Fuzzy attacks, this result is well compared with the lightweight model. 

\subsection{Generalization performance}
\begin{figure}[t]
    \centering
    \includegraphics[width=1\linewidth]{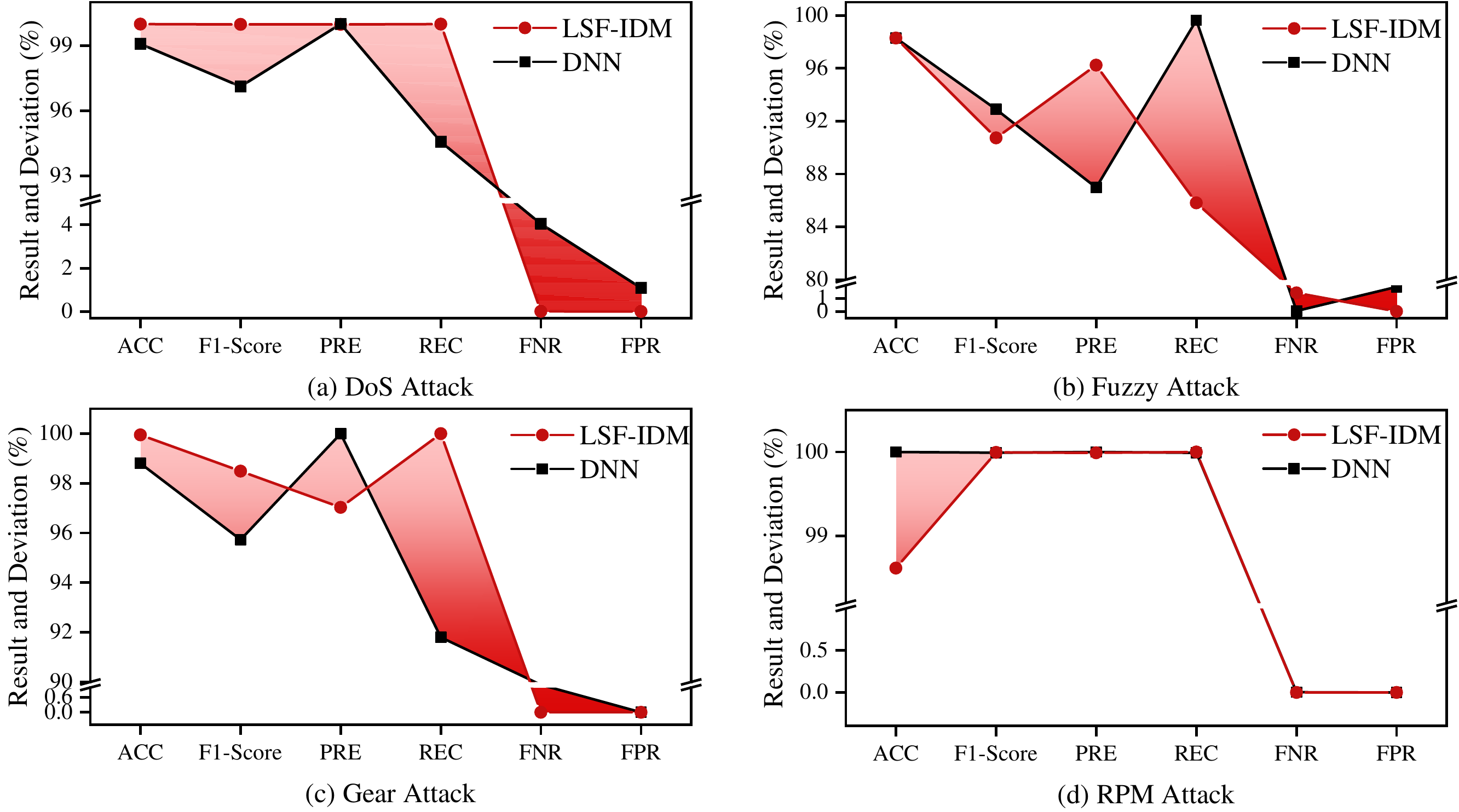}
    \caption{Performance evaluation on each type of attack of the HCRL dataset, where each type from (a) to (d) consists of evaluation performance on six metrics between the DNN and the LSF-IDM. }
    \label{fig8}
\end{figure}
In this subsection, the proposed model performs a generalized performance analysis on a simple model-DNN. The experimental results on the HCRL dataset are shown in Fig.~\ref{fig8}. Overall, it can be seen that when the DNN learns the knowledge that is context-semantic from the teacher model, the performance on the test sets has the presence of instability compared to BiLSTM, attributed to it can only capture simple non-linear features rather than the context information.

Fortunately, the context-semantic information also provides a part of contributes to it, which can be seen in the DoS attack with a higher F1-score and lower false alarm. In the scenario of the RPM attack, the proposed model maintains all performance but has a relative deviation on ACC. We suppose that the context-semantic may noise feature to DNN because of its inherent ability to quickly capture the patterns of RPM attacks. Moreover, we observe that this model more focus on recall in the Gear attack, which contributes to an improvement in F1-score. In contrast, it has a significant decrease in recall of the Fuzzy attack despite concentrating on the improvement of precision. The reason is that fused features analyze complex Fuzzy attacks that are randomly generated, which may form adversarial learning without any advantage for the proposed model. Moreover, we observe that this model more focus on recall in the Gear attack, which contributes to an improvement in F1-score. In contrast, it has a significant decrease in recall of the Fuzzy attack despite concentrating on the improvement of precision. The reason is that fused features analyze complex Fuzzy attacks that are randomly generated, which may form adversarial learning without any advantage for the proposed model. However, it is worth noting that the LSF-IDM in DNN presents an equivalent or lower false alarm on four attacks, which proves the importance of the context-semantic feature directly. On the other hand, we can clarify that the LSF-IDM has the generalization capability even if using a simple model.

\begin{table}[t]
\centering
\caption{Performance difference analysis of LSF-IDM with BERT and DNN }
\label{tab6}
\resizebox{\linewidth}{!}{
\begin{tabular}{llllllll}
\toprule
Attack type                  & Model                                   & ACC/\% & F1-Score/\% & PRE/\% & RE/\% & FNR/\%      & FPR/\%      \\ \hline
\multirow{5}{*}{DoS}   & BERT                                 & 100    & 100         & 100    & 100   & 0           & 0           \\
                        & \multirow{3}{*}{LSF-IDM (DNN)} & \textcolor{green}{-0.01}  & \textcolor{green}{-0.02}       & \textcolor{green}{-0.03}      & \textcolor{green}{-0.02} & \textcolor{green}{+0.01}          & \textcolor{green}{+$2.99\times 10^{-3}$}           \\
                       &                                      & 99.99	&99.98	&99.97	&99.98	&0.01	&$2.99\times 10^{-3}$           \\
                       &                                      & \textcolor{red}{+0.91}     & \textcolor{red}{+2.87}       & \textcolor{red}{-0.03}      & \textcolor{red}{+5.41} & \textcolor{red}{-1.08}       & \textcolor{red}{$-2.99\times 10^{-3}$}           \\
                       & DNN                               & 99.08  & 97.12       & 100    & 94.57 & 1.09        & 0           \\ \midrule
\multirow{5}{*}{Fuzzy} & BERT                                 & 99.98  & 99.98       & 99.97  & 99.98 & 0.01        & $2.99\times10^{-3}$  \\
                       & \multirow{3}{*}{LSF-IDM (DNN)} & \textcolor{green}{-1.69}  & \textcolor{green}{-9.25}      & \textcolor{green}{-3.73}  & \textcolor{green}{-14.17} & \textcolor{green}{+1.41}       & \textcolor{green}{$+1.90\times 10^{-3}$}       \\
                       &                                      & 98.29  & 90.73       & 96.24  & 85.81 & 1.42        & 0.19        \\
                       &                                      & \textcolor{red}{+0}  & \textcolor{red}{-2.16}       & \textcolor{red}{+9.27} & \textcolor{red}{-13.82} & \textcolor{red}{+1.37}       & \textcolor{red}{-1.70}       \\
                       & DNN                               & 98.29  & 92.89       & 86.97  & 99.63 & 0.05        & 1.89        \\ \midrule
\multirow{5}{*}{Gear}  & BERT                                 & 100    & 100         & 100    & 100   & 0           & 0           \\
                       & \multirow{3}{*}{LSF-IDM (DNN)} & \textcolor{green}{-0.05}      & \textcolor{green}{-1.51}           & \textcolor{green}{-2.97}      & \textcolor{green}{0} & \textcolor{green}{0}           & \textcolor{green}{0}           \\
                       &                                      & 99.95    & 98.49       & 97.03    & 100   & 0           & 0           \\
                       &                                      & \textcolor{red}{+1.15}  & \textcolor{red}{+2.76}       & \textcolor{red}{-2.97}  & \textcolor{red}{+8.2} & \textcolor{red}{-1.39}       & \textcolor{red}{0}  \\
                       & DNN                               & 98.80  & 95.73       & 100  & 91.80 & 1.39        & 0   \\ \midrule
\multirow{5}{*}{RPM}   & BERT                                 & 100    & 100         & 100    & 100   & 0           & 0           \\
                       & \multirow{3}{*}{LSF-IDM (DNN)} & \textcolor{green}{-1.38}      & \textcolor{green}{0}           & \textcolor{green}{-0.01}      & \textcolor{green}{0}     & \textcolor{green}{0} & \textcolor{green}{0} \\
                       &                                      & 98.62    & 100         & 99.99    & 100   & 0 & 0 \\
                       &                                      & \textcolor{red}{-1.38}  & \textcolor{red}{+0.01}       & \textcolor{red}{0}      & \textcolor{red}{+0.02} & \textcolor{red}{$-2.21\times 10^{-3}$}     & \textcolor{red}{0}   \\
                       & DNN                               & 100  & 99.99       & 99.99    & 99.98 & $+2.21\times 10^{-3}$        & 0           \\ \bottomrule
\end{tabular}}
\end{table}
From Table~\ref{tab6}, we also present the performance difference analysis for DNN. With respect to the lightweight model, the testing for the ACC metric on the LSF-IDM (DNN) achieves 99.99\% on DoS ($0.91\%\uparrow$), 98.29\% on Fuzzy, and 99.95\% on Gear attack ($1.15\%\uparrow$). Consistent with the previous analysis, it can be sensitive to the RPM attack without the need for redundancy features. In terms of precision, the knowledge transfer behavior is well without any loss on DoS and Gear attacks, resulting in 99.98\%(5.41\%$\uparrow$) and 100\% (8.2\%$\uparrow$) compared to the lightweight model. Although having a relatively weak performance on recall, the F1-score has enough competition with 99.98\% (2.87\%$\uparrow$) and 98.49\% (2.76\%$\uparrow$).
In contrast, the dramatic decrease of recall on the Fuzzy attack makes the F1-score reduce by 2.16\%. 

In particular, we observe that LSF-IDM (DNN) does not generate false positives for DoS and spoofing attacks, benefiting from valuable knowledge instruction. Although improving the identification capacity of the fuzzy attack, the overly sensitive recognition makes for a high FNR.

\subsection{Comparison to the Baselines}
In this subsection, we compared it with four baseline methods, i.e., threshold-based method (OTIDS~\cite{lee2017otids}), self-supervised based~\cite{song2021self}, and LSTM-based~\cite{taylor2016anomaly}, respectively. The proposed evaluation (precision, recall, and F1-score) metrics of different baselines against LSF-IDM are reported in Table~\ref{tab7}. Note that we only compared with some previous lightweight models as the proposed model has an equivalent or proximity amount of parameters with them.

\begin{table}[t]
\caption{Performance Comparison}
\label{tab7}
\centering
\begin{tabular}{lllll}
\hline
Attack Type                     & Model & F1/\% & PRE/\% & RE/\% \\ \hline
\multirow{4}{*}{DoS}   & \cite{lee2017otids}   & 84.88 & 99.90  & 73.80 \\
                         & \cite{song2021self}   & 98.33 & 97.51  & 99.16 \\
                         & \cite{taylor2016anomaly}   & 94.30 & 95.80  & 95.80 \\
                         &  LSF-IDM (Ours)  & \textbf{99.99} & \textbf{100}    & \textbf{99.99} \\ \midrule
\multirow{4}{*}{Fuzzy} &  \cite{lee2017otids}  & 92.80 & 93.30  & \textbf{93.30} \\
                         &  \cite{song2021self}  & 93.05 & 94.45  & 83.45 \\
                         & \cite{taylor2016anomaly}   & 82.20 & \textbf{99.14}  & 70.21 \\
                         &  LSF-IDM (Ours) & \textbf{94.59} & 98.18  & 91.53 \\ \midrule
\multirow{4}{*}{Gear}  & \cite{lee2017otids}   & 91.31 & 91.83  & 90.80 \\
                         & \cite{song2021self}   & 92.61 & 97.68  & 88.03 \\
                         & \cite{taylor2016anomaly}   & 83.42 & 99.83  & 71.65 \\
                         & LSF-IDM (Ours)   & \textbf{100}   & \textbf{99.99}  & \textbf{100}   \\ \midrule
\multirow{4}{*}{RPM}   &  \cite{lee2017otids}  & 91.30 & 91.80  & 90.80 \\
                         &  \cite{song2021self}  & 98.50 & 97.20  & 99.83 \\
                         &  \cite{taylor2016anomaly}  & 83.43 & 99.81  & 71.68 \\
                         & LSF-IDM (Ours)   & \textbf{100}   & \textbf{100}    & \textbf{100}   \\ \bottomrule
\end{tabular}
\end{table}
In Table~\ref{tab7}, we observe that after knowledge distillation, (a) our LSF-IDM achieves higher performance on the DoS attack compared to baseline models. Although the threshold-based (OTIDS) has the proximate capability (99.99\% vs 100\%), the worst recall leads to 84.88\% F1-score with a gap of 15.11\%. The original LSTM-based IDS only gets 94.3\% F1-score, compared to our method (99.99\%), proving the contribution of context-semantic. The core inspiration of Self-supervised based IDS is similar to us learning context information but has an extent deviation of performance with respect to our method. (b) On Spoofing attacks, the result shakes the lead with baseline models, especially the LSTM-based (100\% vs 83.42\% and 100\% vs 83.43\% in F1-score). The reason is that the LSTM-based method could not guarantee a stable performance on recall. (c) The proposed method achieves optimal performance on the Fuzzy attack with a 94.59 F1-score, despite the precision and recall are not outstanding. This reason is attributed to the baseline method can not maintain the detection capability on precision and recall simultaneously.

\section{Conclusion and future work}\label{sec6}
A variety of cyber-attack can be dangerous threats to the vehicular network as a number of interfaces are open to the external environment. However, deep learning-based IDSs are either limited by the computational resources required for complex models or by performance bottlenecks for lightweight models. To this end, we propose LSF-IDM, with lightweight attribution and fused features against malicious traffic using knowledge distillation. We demonstrate the effectiveness of LSF-IDM in the detection of representative attacks. 

For future research, LSF-IDM can be implemented with different pre-trained language models and some state-of-art knowledge distillation methods. Moreover, we hope that the proposed model can be analyzed for how to conduct knowledge transfer with some interpretation works.

\section*{Acknowledgments}
This work was supported in part by the Graduate Research and Innovation Projects of Jiangsu Province under Grant KYCX20\_2862; in part by the China Scholarship Council under Grant 202008690005; the Joint Funds of the National Natural Science Foundation of China (Grant No. U21B2020); Shanghai Science and Technology, China Plan Project (Grant No. 22511104400).

\bibliographystyle{unsrt}  
\bibliography{references}

\end{document}